\documentclass[aps,prx,twocolumn,superscriptaddress]{revtex4-2}
\usepackage{graphicx,SIunits}
\usepackage{bm}     
\usepackage{hyperref} 
\usepackage{dsfont}    
\usepackage{color}
\usepackage{mathtools}
\DeclarePairedDelimiter\ket{|}{\rangle}

\begin{document}
\title{Encoded-Fusion-Based Quantum Computation for High Thresholds with Linear Optics}

\author{Wooyeong Song}
\affiliation{Center for Quantum Information, Korea Institute of Science and Technology (KIST), Seoul 02792, Republic of Korea}

\author{Nuri Kang}
\affiliation{Center for Quantum Information, Korea Institute of Science and Technology (KIST), Seoul 02792, Republic of Korea}
\affiliation{Department of Physics, Korea University, Seoul 02841, Republic of Korea}

\author{Yong-Su Kim}
\affiliation{Center for Quantum Information, Korea Institute of Science and Technology (KIST), Seoul 02792, Republic of Korea}
\affiliation{Division of Nano and Information Technology, KIST School, Korea University of Science and Technology, Seoul 02792, Republic of Korea}

\author{Seung-Woo Lee}
\email{swleego@gmail.com}
\affiliation{Center for Quantum Information, Korea Institute of Science and Technology (KIST), Seoul 02792, Republic of Korea}

\date{\today} 

\begin{abstract}
\noindent 
We propose a fault-tolerant quantum computation scheme in a measurement-based manner with finite-sized entangled resource states and encoded fusion scheme with linear optics. The encoded-fusion is an entangled measurement devised to enhance the fusion success probability in the presence of losses and errors based on a quantum error-correcting code. We apply an encoded-fusion scheme, which can be performed with linear optics and active feedforwards to implement the generalized Shor code, to construct a fault-tolerant network configuration in a three-dimensional Raussendorf-Harrington-Goyal lattice based on the surface code. Numerical simulations show that our scheme allows us to achieve up to 10 times higher loss thresholds than nonencoded fusion approaches with limited numbers of photons used in fusion. Our scheme paves an efficient route toward fault-tolerant quantum computing with finite-sized entangled resource states and linear optics. 

\end{abstract}

\maketitle

Toward scalable quantum computation \cite{Shor96, Fowler09, Fowler12, Raussendorf06, Raussendorf07}, photonic systems have been considered as leading platforms thanks to high-quality sources and detectors, efficient modularity and connectivity, and long decoherence time at room temperature \cite{Slussarenko19, Takeda19, Browne05}. Especially, extremely fast measurements on photons make them suited for measurement-based quantum computing~\cite{Mercedes15, Li15, Herr18, Auger18, Pant19}. 
In measurement-based quantum computing, universal gate operations are realizable via single-qubit measurements applied on entangled resource states prepared offline. However, due to the nondeterministic fusion~\cite{Weinfurter94, Calsamiglia01}--a projective measurement on entangled photons--and loss in photonic platforms, an extensive number of entangled photons are consumed to prepare the resource states for fault-tolerant architectures~\cite{Mercedes15, Li15, Herr18, Auger18, Pant19}. 
 
To circumvent such formidable prerequisites, fusion-based quantum computing (FBQC) was recently proposed~\cite{Bartolucci23, Sahay23, Paesani23}, performed via fusions between constant-sized resource states without extensive entanglement prepared and with stability maintained during the process. 
Its architecture consists of resource states and fusions, which are connected to each other to create a specific network configuration called a fusion network. By constructing a fusion network, a quantum error-correcting (QEC) code can be implemented. For example, surface code is implemented as three-dimensional Raussendorf-Harrington-Goyal (RHG) lattice~\cite{Raussendorf06, Raussendorf07, Fowler09, Fowler12}. 
The details of FBQC are in Ref.~\cite{Bartolucci23}. The fusion thus plays a crucial role in FBQC and its efficiency directly affects the computation performance. However, the fusion success probability is limited by 50\% with linear optics. Moreover, its boost with ancillary entangled photons~\cite{Grice11} turned out to be in a trade-off with the loss tolerance~\cite{Bartolucci23}.
Therefore, fusions in the presence of loss degrade the performance of FBQC significantly, which becomes more crucial when the system size increases, and, as a result, it may be still challenging to build a fault-tolerant architecture in photonic quantum computing platforms. 

In this Letter, we propose a scheme for fault-tolerant quantum computation with finite-sized entangled states and fusions protected by QEC. 
An {\em encoded fusion} is devised to enhance the fusion success probability under loss by QEC. We apply an encoded fusion designed based on $(n,m)$-generalized Shor \cite{Shor95} or parity code \cite{Ralph05, Ewert16, Lee19}, implementable with linear optics and active feedforwards, to construct a RHG lattice. Numerical simulations show that our scheme achieves up to 10 times higher loss thresholds for individual photons than nonencoded fusion approaches \cite{Bartolucci23, Sahay23, Paesani23} with a limited number of photons used per fusion. Specifically, a record-high threshold 14\% is achieved with moderate encoding numbers, e.g.,~$(7,4)$ with single-step feedforward. We also show that when adopting the same encoded resource states, our scheme can reach significantly higher loss thresholds than FBQC~\cite{Bartolucci23} by consuming fewer photons.

Our approach, while motivated from Ref.~\cite{Bartolucci23}, offers a different way toward fault tolerance. The result demonstrates that a concatenation of two QECs, one for the fusion and the other for the network configuration, can dramatically enhance the loss thresholds. A similar approach has been recently introduced in Ref.~\cite{Bell23}. We here focus on RHG lattice and resource states used in Ref.~\cite{Bartolucci23} for direct comparison, but our scheme is not limited to a specific configuration but generally applicable for various architectures and resource states~\cite{Sahay23, Paesani23}.

\begin{figure}[t]
  \includegraphics[width=3.0in]{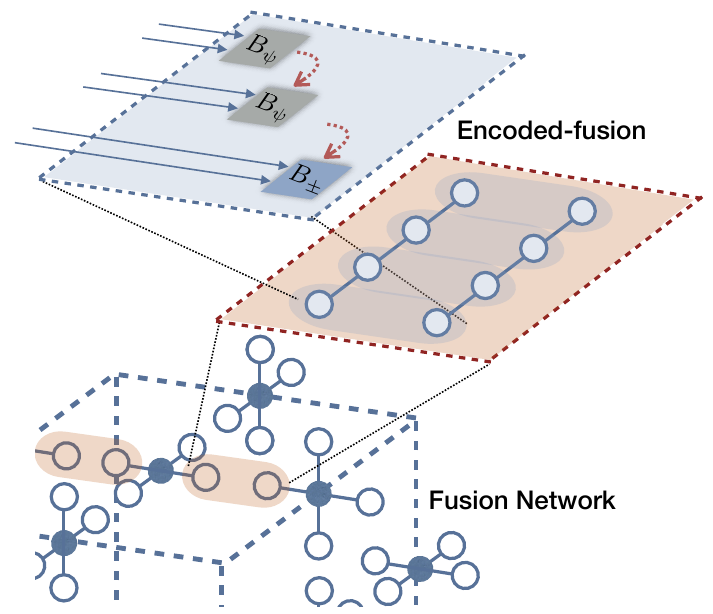}
  \caption{Schematics of EFBQC. In a fusion network, the photons participating in fusions are encoded in a QEC code, and an encoded-fusion protocol is performed actively in a concatenative manner between encoded qubits.}
  \label{scheme}
\end{figure}

\textit{Encoded-fusion-based quantum computation.---}
Let us introduce the encoded-fusion-based quantum computing (EFBQC). In EFBQC, the process to create a fusion network and logical gate operations is conceptually equivalent to FBQC~\cite{Bartolucci23} except that fusions are replaced with encoded fusions. Compared to FBQC, however, EFBQC is aimed more at correcting the errors from resource states, fusion failure, and photon loss by fusion itself, while constructing a fusion network, as illustrated in Fig.~\ref{scheme}. Fusions are applied between resource states to construct a specific fusion network. A fusion network can thus be designed appropriately such that all measurements are projections onto stabilizer states, and corresponding QEC schemes based on the stabilizer formalism can then be applied to achieve the fault tolerance. We here focus on fusion networks in which fusions are the projection onto a particular stabilizer basis, i.e.,~the Bell basis. Such a fusion, called Bell fusion, can be described as $X_1X_2$ and $Z_1Z_2$ measurements on two qubits 1 and 2, whose operators form a stabilizer group $\langle X_1 X_2, Z_1 Z_2 \rangle $. The outcomes of all Bell fusions combine to perform parity checks to enable error correction, e.g., by surface code.

To realize FBQC with linear optics, we should account for two imperfections leading to erasures of measurement outcomes: (i) photon loss, a dominant error source in any photonic platforms, and (ii) the 50\% limit of the success probability of Bell fusion or equivalently Bell-state measurement (BSM) with linear optics. Specifically, a fusion failure can be treated as an erasure of either $X_1X_2$ or $Z_1Z_2$ outcome. Any loss in each fusion causes a complete erasure of outcomes. As a result, such erasures reduce the error tolerance of FBQC significantly, which becomes a crucial factor in building a linear optical scalable architecture. It turned out that boosting the success probability with ancillary entangled photons~\cite{Grice11} increases the rate of erasures and eventually harms the loss tolerance of FBQC~\cite{Bartolucci23}. 

On the other hand, in EFBQC, encoded fusions play a role logically as $X_1X_2$ and $Z_1Z_2$ on two encoded qubits 1 and 2 of loss so that all events of erasures of $X_1X_2$ and $Z_1Z_2$ can be suppressed. Therefore, all fusion outcomes are consistent with the resource state stabilizers, and, in principle, error correction in fusion network exhibits the maximum performance of the fault tolerance (see Appendix~\ref{a:FBQC} for details). 
An encoded fusion can be implemented by performing multiple linear-optic BSMs consecutively with a QEC protocol that enables increasing the success probability even in the presence of photon loss, as we introduce in the following.

\textit{Encoded fusion with linear optics.---} 
Several schemes have been proposed to overcome the 50\% limit of the fusion success probability with linear optics by using ancillary entangled photons~\cite{Grice11, Ewert14}, squeezing~\cite{Zaidi13}, and Greenberger-Horne-Zeilinger (GHZ) encoding~\cite{Lee15,Lee15a}. However, employing a large number of photons in fusion generally is at a higher risk of photon loss, which offsets an advantage and eventually is in a trade-off with the loss threshold of FBQC~\cite{Bartolucci23}. In contrast, an encoding only against photon loss does not solve the problem induced by the low efficiency of fusion. Therefore, it is essential to enhance the success probability of fusion while suppressing the effects of photon loss that may occur in the fusion and resource state preparation.

We introduce a method to enhance both the fusion success probability and loss tolerance by a QEC protocol with linear optics. Consider the $(n, m)$-Shor or parity code~\cite{Ralph05} with dual-rail qubits as a representative example. We define the logical basis as $|0_L\rangle = |+^{(m)}\rangle^{\otimes n} $ and $|1_L\rangle = |-^{(m)}\rangle^{\otimes n} $, where $|\pm^{(m)}\rangle = (|H\rangle^{\otimes m} \pm |V\rangle^{\otimes m})/\sqrt{2}$ consists of $n$ blocks each of which includes $m$ photons in $|\pm\rangle $ state. Interestingly, the encoded Bell states $|\Psi^\pm\rangle = |0_L\rangle|1_L\rangle \pm |1_L\rangle |0_L\rangle $ and $|\Phi^\pm\rangle = |0_L\rangle|0_L\rangle \pm |1_L\rangle |1_L\rangle $ can be decomposed into $n$ number of block-level Bell states, each of which in turn is decomposed into $m$ number of photonic Bell states. Appendix~\ref{a:EBS} includes a detailed description of the decomposition. While a linear-optic BSM can discriminate only two out of the four Bell states, such characteristics of the encoded states make it possible to logically distinguish the Bell states by a series of $n \times m$ linear-optic BSMs with much higher efficiencies.

We now sketch the encoded-fusion protocol based on linear optics and active feedforwards (details in Appendix~\ref{a:EBS}). In physical qubit level, we use three types of linear-optic BSMs discriminating  $|\psi^+\rangle / |\psi^-\rangle$, $|\psi^+\rangle / |\phi^+\rangle$ and $|\psi^-\rangle / |\phi^-\rangle$ deterministically, denoted as $B_\psi$, $B_+$ and $B_-$, respectively. Note that BSM can be implemented by basic linear-optical elements such as polarizing beam splitters, wave plates and photon detectors, which can discriminate only two out of the four Bell states. The type can be easily changed by simply rotating wave plates on the inputs of polarizing beam splitters. The protocol is as follows:

(i) In each block, $B_{\psi}$ is applied on each pair of photons randomly selected from distinct encoded qubits. Repeat until $B_{\psi}$ succeeds, detects a loss, or consecutively fails $j\leq m-1$ times (a predetermined optimized number).

(ii) $B_{+}$ or $B_{-}$ is applied on the remaining photon pairs, if any $ B_{\psi}$ succeeded with the result $\ket{\psi^+}$ and $\ket{\psi^-}$, respectively. For loss detection and $j$-times failure, $B_{+}$ or $B_{-}$ is randomly selected. 

(iii) Total $n$ times of block-level protocols, (i) and (ii), are performed independently. 

In each block, the sign ($\pm$) is identified by any success of $B_{\psi}$, and the letter ($\psi,\phi$) is also identified based on the results of all $B_{\pm}$ performed on remaining photon pairs. So, full discrimination is possible unless loss occurs, and at least the sign can be identified with any single success of $ B_{\psi}$ or $B_{\pm}$. We denote the full discrimination and failure probability as $p_s$ and $p_f$, respectively. The probability of only sign discrimination is then $1-p_f-p_s$. 

By collecting all the results of independently performed $n$-times block-level protocols, the logical result is determined. The letter is the same with any determined block-level result. The sign can be identified by counting the number of minus ($-$) signs from block-level results. As a result, the success probability of encoded fusion based on $(n,m)$-Shor code is obtained as $P_s(\eta) = (1-p_f)^n - (1-p_s -p_f)^n$ with a given loss rate $\eta$ per photon, which becomes $1-2^{-mn}$ when no loss occurs. Note that, in contrast to the boost scheme with ancillary entangled photons \cite{Grice11}, the encoded fusion can succeed with arbitrarily high rates with a moderate encoding number $(n,m)$ in the presence of photon loss. See Appendix~\ref{a:EBS} for details.

\begin{figure}[t] 
  \includegraphics[width=3.3in]{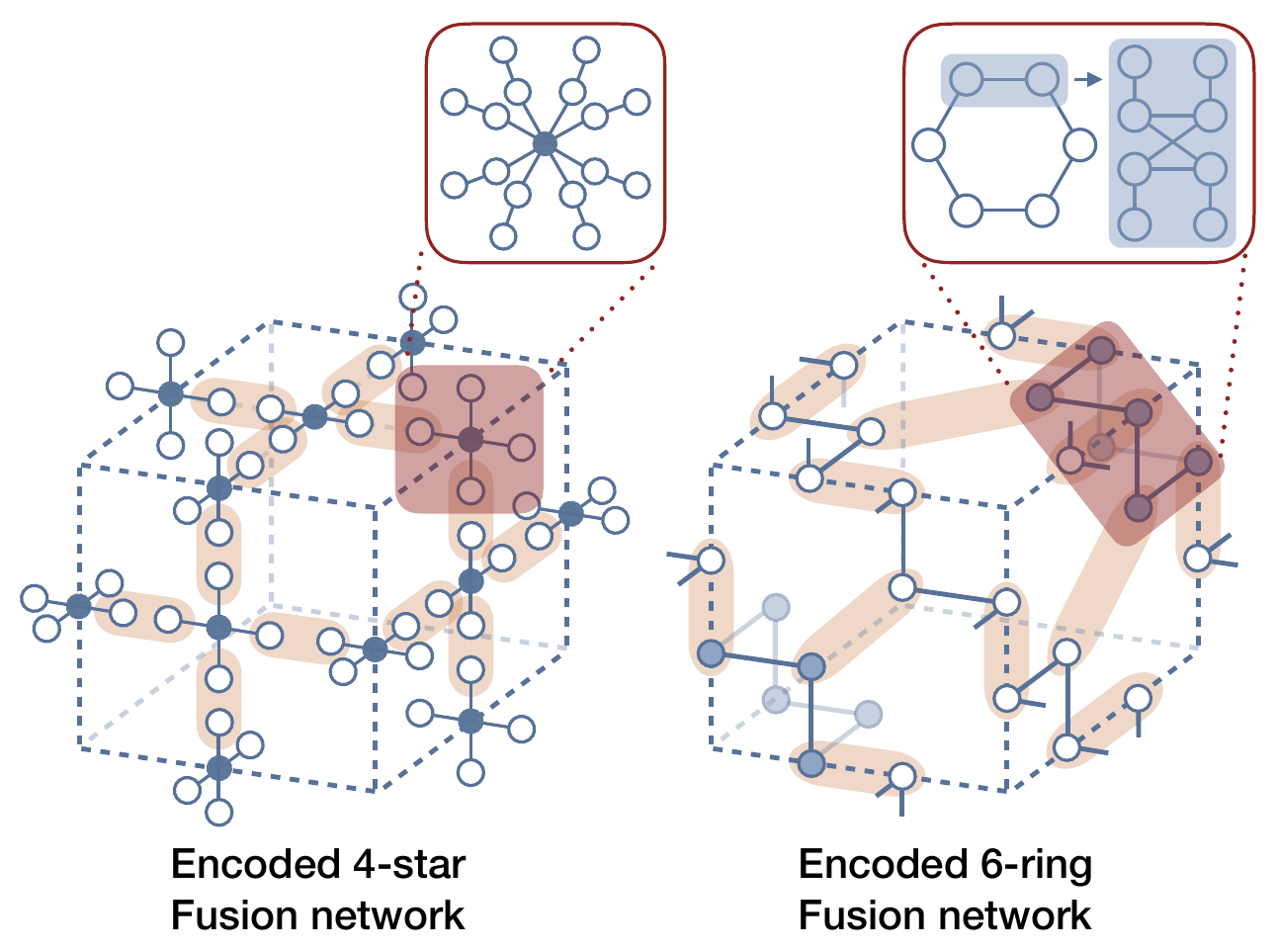}
 \caption{For the direct comparison with FBQC~\cite{Bartolucci23}, we apply our scheme to the networks in RHG lattice fabricated with the encoded 4-star and encoded 6-ring resource states. The insets illustrate the simplest example with $(n,m)=(2,2)$. The structure and encoded resource states are the same with FBQC \cite{Bartolucci23}, while the fusions are replaced with the encoded fusions in orange ovals.}
  \label{fig:EFNexample}
\end{figure}

\textit{Encoded-fusion networks and resource states.---}
A fusion network is constructed to implement a foliated QEC code \cite{Bolt16}. A standard approach implementing the surface code leads to form a three-dimensional RHG lattice~\cite{Raussendorf06, Raussendorf07, Fowler09, Fowler12}. A variation of surface code for biased noises, i.e.,~\textit{XZZX} code \cite{Ataides21, Claes23}, can be implemented by constructing a \textit{XZZX} lattice fusion network~\cite{Sahay23}. The aforementioned lattice models were shown to be fabricated by employing 4-star and 6-ring shape resource states introduced in Ref.~\cite{Bartolucci23}. Linear cluster states can also be used as resource states \cite{Paesani23} to create a foliated Floquet color code architecture~\cite{Hastings21, Davydova23}. We here focus on RHG lattice structures for the direct comparison with FBQC \cite{Bartolucci23}. 

The process to form a RHG lattice and the corresponding resource states are logically equivalent to FBQC. The encoded-fusion networks in RHG lattice can thus be constructed by applying encoded fusions on the encoded 4-star or 6-ring resource states as illustrated in Fig.~\ref{fig:EFNexample}. However, not only the resource states but also the fusion schemes are here reformulated as encoded forms, e.g.,~by $(n,m)$-Shor code in the current model. The encoded 4-star and 6-ring resource states have the forms obtained by replacing all the qubits participating in fusion  with encoded qubits. As simplest examples, $(2,2)$ encoded 4-star and 6-ring resource states are illustrated in the inset of Fig.~\ref{fig:EFNexample}. Such encoded resource states in arbitrary encoding numbers can be generated straightforwardly by fusing entangled resource states such as GHZ states \cite{Mercedes15}. 
For example, the encoded 4-star resource state based on $(n,m)$-Shor code is composed of $4\times(n \times m)$ photonic qubits, and can be generated by fusing $4n$-GHZ state and $4 \times n$ number of $(m+1)$-GHZ states. The generation schemes of encoded resource states are elaborated in Appendix~\ref{a:RS}. Note that arbitrary $n$-GHZ states can be built from 3-GHZ states that are readily available in current photonic technologies~\cite{Schwartz16, Istrati20, Uppu21, Lu21, Bartolucci21_1, Economou22, Thomas22, Cogan23}. Once the resource states are prepared with an encoding number $(n,m)$, the encoded fusions with the same $(n,m)$ are correspondingly applied.

\textit{Thresholds of encoded-fusion networks.---}
The performance of fusion networks can be analyzed with two error models: (i) hardware-agnostic error model with the erasure rate $P_{\text{erasure}}$ and the measurement (flipped) error rate $P_{\text{error}}$, and (ii) linear-optical error model with the fusion success rate $P_s(\eta)$ and the loss rate $\eta$ for individual photons. The thresholds of FBQC was analyzed in Ref.~\cite{Bartolucci23}, in which the correctable regions of two parameters $P_{\text{erasure}}$ and $P_{\text{error}}$ were evaluated by Monte Carlo simulation, e.g., yielding $P_{\text{erasure}}$ thresholds $6.90\%$ for 4-star and $11.98\%$ for 6-ring fusion networks when no measurement error occurs. Photon loss thresholds under the linear-optical error model were then estimated as about $0.25\%$ and $0.78\%$ per individual photon for 4-star and 6-ring fusion networks, respectively~\cite{Bartolucci23}, assuming boosted fusion success probabilities with ancillary entangled photons \cite{Grice11}. It was also shown that FBQC using encoded qubits in $(2,2)$-Shor code with boosting can achieve a higher threshold, e.g.,~$2.7\%$ for 6-ring fusion network~\cite{Bartolucci23}.

\begin{figure}[t]
  \includegraphics[width=3.2in]{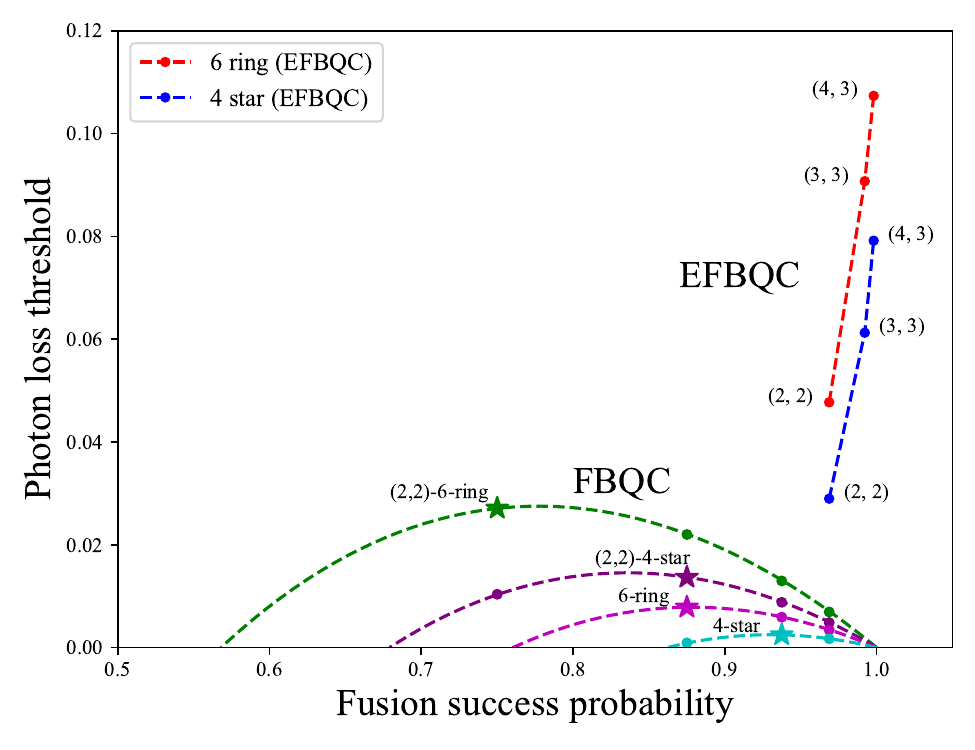}
  \caption{Photon loss thresholds for different fusion success probabilities $P_s(\eta)$. The green, purple, magenta and cyan curves show the results of FBQC in Ref.~\cite{Bartolucci23}: the dots on the curves represent the cases when the fusion success probability is boosted with different numbers of additional entangled photons, e.g.,~$P_s(\eta)=0.75$ with additional 2 photons per physical fusion (additional $2\times4 = 8$) for $(2,2)$, and the star shows the maximum value among them. The blue and red dots represent the loss thresholds of EFBQC with different $(n,m)=(2,2), (3,3), (4,3)$ based on encoded 4-star and 6-ring fusion networks, respectively.}
  \label{PLPS}
\end{figure}

\begin{figure}[t]
  \includegraphics[width=3.2in]{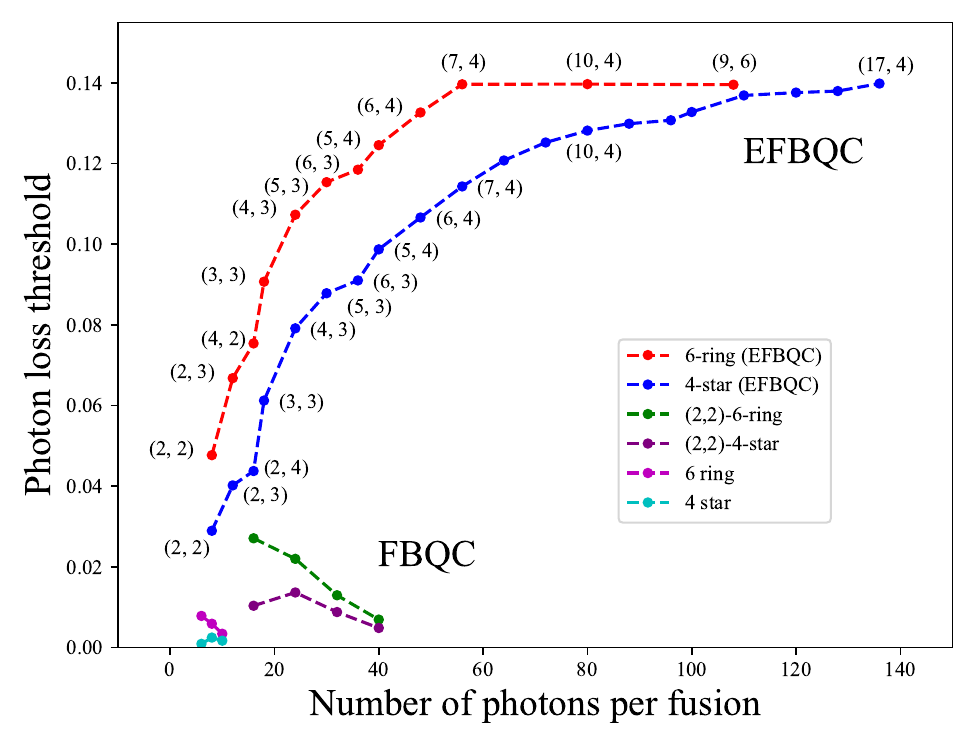}
  \caption{Photon loss thresholds for the total number of photons used per fusion. The thresholds of EFBQC are maximized by optimizing the encoded-fusion protocol for a given encoding number $(n,m)$. The threshold for EFBQC generally gets higher when increasing the number of photons used per fusion, while the threshold for FBQC boosted with ancillary entangled photons decreases. EFBQCs for encoded 4-star and 6-ring resource states, respectively, yield $11.44\%$ and $13.97\%$ when $(n,m)=(7,4)$, and both arbitrarily reach up to $14\%$ as increasing $(n,m)$.}
  \label{PLPN}
\end{figure}

Let us now examine the loss thresholds of EFBQC. We employ the same hardware-agnostic model for the direct comparison with FBQC, resulting in the same correctable regions of $P_{\text{erasure}}$ and $P_{\text{error}}$. We can then estimate the loss thresholds of encoded-fusion networks based on the linear-optical error model characterized by $P_s(\eta)$ and $\eta$ by evaluating $P_{\text{erasure}}$ and $P_{\text{error}}$. We plot the thresholds of EFBQC against the fusion success probability in Fig.~\ref{PLPS} and by changing the total number of photons used per fusion in Fig.~\ref{PLPN} with different fusion encoding numbers $(n,m)$. For comparison, we also plot the thresholds of FBQC obtained in Ref.~\cite{Bartolucci23}.
  
Figure~\ref{PLPS} shows that EFBQC yields much higher loss thresholds and fusion success probabilities than FBQC. The thresholds of FBQC are maximized under a limited fusion success probability and any further boosting degrades these~\cite{Bartolucci23}. This implies that additional use of photons increases the risk of loss in FBQC so that the fusion success probability is in a trade-off with the threshold. On the other hand, in EFBQC, the loss thresholds can be improved together with the fusion success probability. Our results show that the proposed encoded-fusion scheme can enhance the success probability by increasing the encoding number $(n,m)$ while suppressing the effect of loss simultaneously, so that the loss thresholds of EFBQC can be dramatically improved. A loss threshold $4.8\%$ per photon is achieved with $(2,2)$ encoded 6-ring resource states in EFBQC, which is almost doubled from $2.7\%$ obtained with the same resource states and additional entangled photons for boosting in FBQC~\cite{Bartolucci23}, notably by consuming fewer photons and adding only a two-step linear-optical process ($j=1$). See Appendix~\ref{a:RO} for the comparison of resource overheads.

In Fig.~\ref{PLPN}, we plot the loss thresholds of EFBQC numerically maximized in our protocol for given $(n,m)$, and compare the results with FBQC by changing the total number of photons used per fusion. See Appendix~\ref{a:Ana} for the optimized protocol.
It exhibits that, with a fixed number of photons in fusion, EFBQC can achieve much higher thresholds than FBQC. Remarkably, the attained loss thresholds of EFBQC are about 10 times higher than nonencoded and about 5 times higher than encoded FBQCs that were estimated in previous works~\cite{Bartolucci23,Sahay23,Paesani23}. 
Specifically, EFBQCs with $(7,4)$ encoded 4-star and  6-ring fusion networks, respectively, reach $11.44\%$ and $13.97\%$ loss thresholds per photon. This implies that a moderate number of additional photons used in fusion can substantially enhance the loss thresholds by our scheme. Note that both 4-star and 6-ring encoded-fusion networks can reach arbitrarily up to $14\%$ by increasing the encoding number $(n,m)$. Such a maximum threshold may be the characteristic of current choices of concatenated QEC codes, i.e., generalized Shor and surface code, and thus possibly can be enhanced further with other codes~\cite{Sahay23,Paesani23}.

\textit{Remarks.---} 
We have proposed a fault-tolerant quantum computation scheme performed in a measurement-based manner with finite-sized entangled resource states and encoded fusions. In contrast to FBQC schemes~\cite{Bartolucci23, Sahay23, Paesani23}, two different QECs, one for the fusion and the other for the network configuration, are used concatenatively in EFBQC. The encoded fusion is devised to correct photon loss, fusion failure, and resource errors within the fusion process by implementing a QEC code. Moreover, an encoded fusion with $(n,m)$-Shor code is shown to be efficiently implementable with linear optics and active feedforwards only. We have applied the encoded fusion to construct a fusion network in RHG lattice. By numerical simulations, we have demonstrated that our scheme improves the loss thresholds up to 10 times higher than nonencoded fusion approaches~\cite{Bartolucci23, Sahay23, Paesani23}, and allows us to attain $\sim$14\% loss thresholds per individual photon, which is to our knowledge, a record-high threshold among recent achievements in photonic quantum computing platforms~\cite{Bartolucci23, Sahay23, Paesani23, Lee23}. We have also shown that EFBQC outperforms FBQC with respect to the attainable thresholds by consuming the same number of photons.

We found that Bell \textit{et al.}~\cite{Bell23} have similarly studied encoding for fusion to improve thresholds over FBQC~\cite{Bartolucci23}; 10.5\% is achieved by encoding into a 10-qubit graph code with an adaptive protocol, which is comparable to our results with $(4,3)$ encoded (12-qubit) case, being lower, and higher than $(3,3)$ encoded (9-qubit) case, while our scheme enhances the threshold further up to 14\% by increasing the encoding size. Such an optimal graph state can be searched by an exhaustive search method priorly for a given encoding size~\cite{Bell23}, while applying our scheme for arbitrary high $(n,m)$ is straightforward with the same protocol. Despite being developed independently using different codes and protocols, both schemes provide a common alternative way toward fault tolerance to overcome the limit of standard FBQC. See also Ref.~\cite{Pankovich24} in which high thresholds have been achieved using GHZ-state measurements.

Our scheme can be implemented by linear optics with few-step feedforwards, which is efficiently realizable within current technologies \cite{Prevedel07, Bartolucci21_2}. By simply adding one more step of linear-optical process ($j=1$), EFBQC almost doubles the threshold of $(2,2)$-Shor code encoded 6-ring network estimated in FBQC~\cite{Bartolucci23}. Moreover, numerical optimization shows that only one or two additional steps with a moderate number of photons in encoding, e.g.,~$(7,4)$ with single-step feedforward ($j=1$) can yield loss thresholds up to 14\%. All required encoded resource states are producible with available entangled photon sources~\cite{Schwartz16, Istrati20, Uppu21, Lu21, Bartolucci21_1, Economou22, Thomas22, Cogan23}. Our scheme is thus readily implementable within current and near-term photonic platforms. 

Finally, we note that our approach is not limited to any specific configuration or code, and generally applicable for various architectures by e.g.,~\textit{XZZX} surface~\cite{Sahay23} and Floquet color code~\cite{Paesani23}, and resource states, e.g.,~linear cluster states~\cite{Paesani23}. Developing encoded-fusion protocols with other QECs \cite{Bell23, Bombin23} would be also valuable as next step of research.

This research was funded by Korea Institute of Science and Technology (2E32941) and National Research Foundation of Korea (2022M3K4A1094774).

\clearpage

\appendix

\section{Encoded-fusion and fusion based quantum computation}
\label{a:FBQC}

In this section, we introduce the encode-fusion based quantum computing (EFBQC) by also reviewing the fusion-based quantum computing (FBQC) introduced in Ref.~\cite{Bartolucci23}. 
FBQC is performed by generating a network configuration called \textit{fusion network} that shows a structure with fixed-sized resource states and fusion measurements to be applied on the resource states. 
A fusion is a projective measurement that entangles the remaining qubits, excluding the two qubits that participate in the fusion, which were chosen from two distinct resource states. We note that the process of creating a fusion network, performing gate operations and  fault-tolerance schemes based on all fusion outcomes in EFBQC are logically equivalent to FBQC. In EFBQC, the fusion is replaced with the encoded-fusion, which logically plays the same role as the fusion in FBQC. However, EFBQC is devised to be more aimed at correcting the errors from resource states, fusion failures and photon losses in the fusion process itself by an independent error correction  so that it can reduce the burden of the outer error correction in a fault-tolerance fusion network such as erasures of fusion outcomes. 

\subsection{Fusion network}
\label{a:FN}

Both FBQC and EFBQC employ a fusion to construct a specific fusion network. A fusion network is a configuration that specifies an arrangement of resource states and fusion measurements to be performed. However, note that it does not specify the order of operations, nor mean that all parts of the fusion network should exist simultaneously. 
To implement an error correction scheme for fault-tolerance, a fusion network is designed such that all resource states are stabilizer states and measurements are projections onto stabilizer states. For example, in order to implement the surface code, a fusion network forms a three-dimensional Raussendorf-Harrington-Goyal (RHG) lattice~\cite{Bartolucci23}. The \textit{XZZX} code devised for biased noises can be implemented by fabricating a \textit{XZZX} lattice fusion network~\cite{Sahay23}. The Floquet color code can be  implemented by forming a three-dimensional Floquet color code architecture~\cite{Paesani23}. A corresponding error correction scheme then can be executed based on stabilizer formalism to achieve the fault-tolerance.
We here focus on the projective measurement onto the Bell state, called Bell fusion. While constructing a larger-sized entangled states, a Bell fusion provides stabilizer outputs for the quantum error correction. The Bell fusion can be described as $X_1X_2$, $Z_1Z_2$ measurements on two qubits 1 and 2, and their operators form a stabilizer group $\langle X_1 X_2, Z_1 Z_2 \rangle $. All the outcomes of Bell fusions combine to perform parity checks for a fault tolerance scheme. 

\begin{figure}[b]
  \includegraphics[width=3.3in]{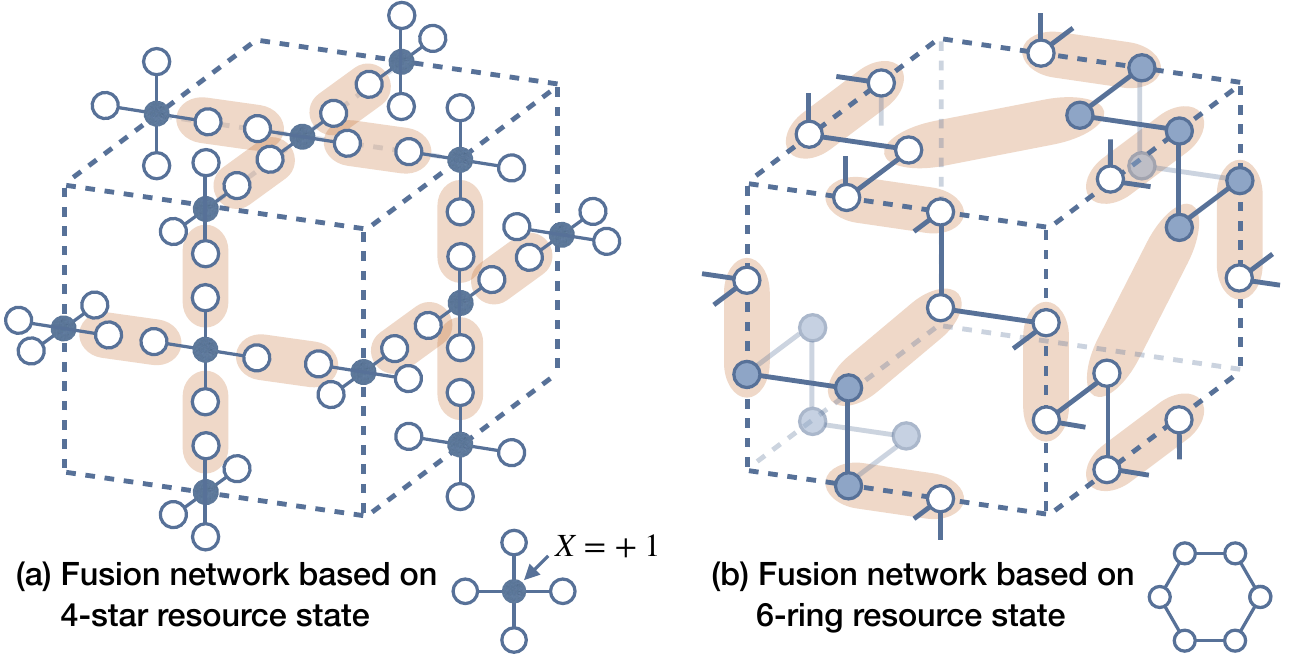}
  \caption{The unit cubic cell of fusion networks constructed based on (a) 4-star resource state and (b) 6-ring resource states~\cite{Bartolucci23}. Orange ovals indicate the fusions. For (b), only colored qubits are in an unit cell, and white qubits represents the qubits in the neighboring cells. In EFBQC, the qubits participating in the fusion and the fusion operations are replaced with the encoded-qubits and -fusions, respectively.}
  \label{FTFN}
\end{figure}

The stabilizer output of a fault-tolerant fusion network can be mainly separated into two subgroups: (i) the output stabilizers that become the logical operators of the system, and (ii) the check operators that provide redundancy in the fusion outcomes. Without errors, the generators in the check operators group has positive eigenvalues. In the presence of a (detectable) error, the eigenvalues of one or more check operators become negative. The value of all check operators is referred to as the syndrome. A fault-tolerant fusion network has a specific geometry according the resource state to leverage the redundancy. 

Let us first consider a fusion network based on 4-star resource states in the form of the RHG lattice. Note that the central qubits of each resource states are measured on a specific basis ($X$) in advance and start the computation with a specific value ($+1$).
As illustrated in Fig.~\ref{FTFN} (a), the unit cubic cell of the lattice based on 4-star resource states is composed of 18 resource states (6 for faces, 12 for edges) and each face of cell has 5 resource states. 
On the face, the resource states are placed in parallel to the face, while the resource states on the edge are aligned perpendicular to the edge. 
Fusion measurements are applied on neighboring qubits from adjacent resource states, i.e., one is from the face and the other is from the edge, so that each face of the unit cell includes 4 fusions. More specifically, each qubit (except the central one) of central resource state located at each face of the unit cell participates in fusions with one qubit of each of the neighboring resource states at edges. One of the remaining qubits from the resource states at each edge participates in the fusion of the adjacent surface of the unit cell, while the other two qubits are used for fusions in the adjacent unit cell. As a result, each unit cell has total 24 fusion measurement outcomes to be used for cell parity check.

A fusion network based on 6-ring resource states in RHG lattice is illustrated in Fig.~\ref{FTFN} (b). Only two resource states are contained in each unit cell. Each resource state is placed on three surfaces that create two diagonally opposite vertices of the cubic unit cell.
Three out of six qubits in one resource state are thus placed on each of three surfaces, and the other three qubits are placed on each edge between the surfaces. 
The fusion network based on 6-ring requires smaller number of fusion measurements to build the unit cell of a fault-tolerant fusion network. Total 18 fusion outcomes are used for cell parity check, which is less than the number of outcomes 24 in 4-star based fusion network. The smaller number of fusion outcomes are less influenced by the fusion imperfections so that it leads to higher loss thresholds.

\subsection{Fusion network in photonic platforms}

Both FBQC and EFBQC frameworks do not restrict the hardware for their implementations, but we here focus on photonic platforms. Let us consider a fusion network constructed based on the fusion performed by linear optics. The qubits can then be defined as dual-rail encoding, i.e., two orthogonal photonic modes. The fusion on such qubits can be easily performed based on linear optics, called linear-optic Bell fusion or Bell-state measurement (BSM), by employing basic linear optical elements such as beam-splitters, wave plates and photo-detectors. However, for the realization in photonic platforms and linear optics, two major imperfections should be accounted for: 

(i) photon loss, a dominant source of errors in all photonic platforms 

(ii) 50\% limit of the fusion success probability restricted by linear optics

In a fault-tolerant fusion network, a fusion failure is treated as an erasure of either $X_1X_2$ or $Z_1Z_2$ outcome. If any single or more photons are lost among the photons used in fusion measurements, the measurement outcomes are totally erased, i.e., `erasure' outcome. Therefore, in FBQC, the two imperfections (i) and (ii) can produce `erasures' frequently, as a result, leading to the reduction of error tolerance. 
In previous works, imperfection (ii), i.e.,~the low success probability has been dealt with a boost method by applying ancillary entangled photons~\cite{Grice11}. However, as increasing the success probability, it also increase the risk of imperfection (i), i.e.,~photon loss so that the fusion success probability is eventually in a trade-off relation with the loss tolerance as pointed out in~\cite{Bartolucci23}. Therefore, within previous approaches~\cite{Bartolucci23}, the realization of FBQC in a linear optical architecture may be still challenging. On the other hand, in EFBQC, the encoded-Bell fusion plays a role as $X_1X_2$ and $Z_1Z_2$ measurements on two encoded logical qubits 1 and 2 aiming to suppress all erasure outcomes of $X_1X_2$ and $Z_1Z_2$. All fusion outcomes are then consistent with the resource state stabilizers so that, in principle, the error correction scheme through constructing a fusion network can exhibit the maximum fault-tolerance.


Both FBQC and EFBQC are designed to be executed in a measurement-based manner through a network of resource states and fusion measurements. Therefore, a physical device to generate the resource states to be used in a fusion network can be modeled as a resource state generator (RSG).
RSG can be used as an module component in FBQC or EFBQC architecture, which generates resource states repetitively to send them to the fusion location via waveguides or fibers directly. Note that photonic platforms have high-quality sources and detectors, and their efficiency of modularity and connectivity are very high. Therefore, such an modular photonic architecture can significantly reduce the operational depth, and in turn provides an advantage for scalability. 

A fusion routing determines the connectivity and time-ordering of operations, essentially deciding which resource states will be correlated. This can be achieved through the spatial and temporal configuration of fibers or waveguides, which serve as physical representations of the connections in the fusion network. 
It is worth noting that fusion network routers can also be used as passive memories with a fixed delay to create temporal correlations. The qubits in one time cycle can be then fused with the qubits in arbitrary later time cycle. This makes it possible to create a larger fusion network even with a small number of RSGs by fusing qubits generated from RSG to the qubits from the same RSG in a different time cycle at the expense of time~\cite{Bombin21_A}. A detailed example of physical layout of such a module architecture is illustrated in Ref.~\cite{Bartolucci23}.

In EFBQC, a fusion device is also designed with multiple settings of modules which enable to implement a specific quantum error-correcting protocol, while in FBQC it is performed by a standard linear-optic way of BSMs for basis configuration. In our approach, the encoded-fusion device can be designed in a concatenative way by basic linear optical components and detectors as well as additional setups for active feed-forwards. The detailed structure and layout as well as the protocol of the encoded-fusion will be introduced in the following section in Appendix~\ref{a:EF}. In EFBQC, RSGs also produce the encoded-resource states, whose forms are elaborated in Appendix~\ref{a:RS}.
Finally, we note that the size of the computation of FBQC and EFBQC being performed is not determined by the size of resource states but by the size of the fusion network and the number of resource states~\cite{Bartolucci23}.

\section{Encoded-fusion}
\label{a:EF}
In this section, we first explain the Bell fusion in stabilizer formalism to be used in a fault-tolerant fusion network. The encoded-Bell fusion can play a same role as the logical Bell fusion in stabilizer formalism, while it can be protected from photon loss by an independent error-correcting code and also beat the 50\% limit of success probability with linear optics. We then introduce an encoded-Bell fusion scheme for implementing $(n,m)$-generalized Shor code with linear optics and active feed-forwards, which can be used for developing fault-tolerant EFBQC.

\subsection{Bell fusion in stabilizer formalism}
We first review and discuss the Bell fusion within the stabilizer formalism. By fusing two qubits from two distinct resource states, we can create a larger entangled state to form a fusion network. To examine a stabilizer fusion network, we describe the fusion process by stabilizer formalism and assume that all resource states are stabilizer states and fusion measurements are the projection onto the stabilizer basis, i,.e.~Bell basis here.

\begin{figure}[b]
  \includegraphics[width=3.0in]{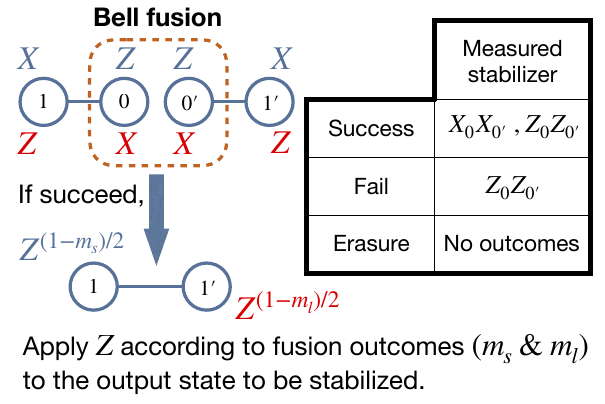}
  \caption{An example of Bell fusion between two graph states. The first 2-qubit state is stabilized by $X_0Z_1$ and $Z_0X_1$, and the second 2-qubit state is stabilized by $X_{0'}Z_{1'}$ and $Z_{0'}X_{1'}$. Here, the fusion is assumed to be performed by typical Type-II fusion with linear optics. The fusion applied between qubit $0$ and $0'$ is equivalent to the measurement on the stabilizer basis $XX$ and $ZZ$. We denote the outcome of $XX$ and $ZZ$ as $m_l$ and $m_s$, respectively. The stabilizer generator after the fusion is determined by its remaining stabilizer generators which are not participating in fusion, together with fusion outcomes $m_l$ and $m_s$. To make the state after fusion stabilized, additional $Z$ operator must be applied according to the corresponding fusion outcome. The table beside represents possible outcomes from Bell fusion. When successful, we can obtain both measurement outcomes. Even if it fails, we can still obtain one outcome, $ZZ$. However, an erasure event would result in no measurement outcome at all.}
  \label{fusions}
\end{figure}

As a simple example, let us consider two distinct graph states having 2 qubits, to see how the new graph state is constructed by fusing them. The two graph states are stabilized by $X_{0}Z_{1}$, $Z_{0}X_{1}$ and $X_{0'}Z_{1'}$, $Z_{0'}X_{1'}$, respectively, as illustrated in Fig.~\ref{fusions}. Specifically, the case that Bell fusion is applied to $0$ and $0'$ qubits as shown in Fig.~\ref{fusions} is equivalent to the measurement on them in the basis $X_0X_{0'}$ and $Z_0Z_{0'}$. If the fusion succeeds and returns $+1$ eigenvalue, the remaining qubits $\{1,1' \}$ are stabilized by $Z_1Z_{1'}$ and $X_1X_{1'}$, corresponding to a 2-qubit graph state. The signs of stabilizers are determined by fusion measurement outcome $m_s$ and $m_l$, so that the output stabilizer generators can be written by 
\begin{eqnarray}
	\langle m_s X_{1}X_{1'}, m_l Z_{1}Z_{1'}  \rangle,
\label{Eq_sg}
\end{eqnarray}
where $m_l$ and $m_s$ are the outcomes of fusion $\langle XX \rangle $ and $\langle ZZ \rangle$, respectively, and $m_s, m_l \in \{\pm1\}$. The obtained graph state after fusion should be corrected by additional Pauli-$Z$ operators $Z^{(1-m_{l(s)})/2}$ as mentioned in Fig.~\ref{fusions}. As a result, we can get new quantum correlations on the remaining qubits together with the signs from fusion measurement outcomes.

A Bell fusion, i.e.,~a projecting measurement onto the stabilizer basis $X_1 X_2$ and $Z_1 Z_2$, with liner optics is inherently non-deterministic process. It gives us one of the three possible events: (i) the fusion succeeds with probability $p_s=1-p_f$, in which the qubits are measured in the $X_1 X_2$ and $Z_1 Z_2$. (ii) the fusion fails with probability $p_f$, which can be described as separable single-qubit measurements $Z_1I_2, I_1Z_2$. In this case, one out of the two desired outcomes $Z_1Z_2$ can be measured, whereas the other $X_1 X_2$ information is completely erased. A fusion failure can thus be regarded as an incomplete BSM in which one of the two measurement outcomes is erased~\cite{Bartolucci23}. Such a biased characteristic of fusion failure can be applied to use a quantum error-correcting code for biased noise to form a specific fusion network structure~\cite{Sahay23,Bombin23_A}. (iii) The last case is fusion erasure, where neither of the stabilizer outcomes  is properly measured. As discussed in the previous section, the two major imperfections in photonic platforms with linear optics, i.e., photon loss and the 50\% limit of the fusion success probability cause the erasure cases by (ii) and (iii). In the following subsection, we will introduce how we can handle both imperfections together to reduce the erasure cases by (ii) and (iii) in fusion outcomes. 

\subsection{Encoded-Bell fusion with linear optics }
\label{a:EBS}

As already notified in previous sections, two issues that we aim to solve are photon loss and the fusion failure. We stress that they should be dealt with at the same time: photon loss and fusion failure. We here propose a linear-optical scheme to implement quantum error correction code in fusion process to increase both the success probability and loss-tolerance simultaneously. Additional photons used in the fusion process do not increase the risk of photon loss but only contribute to enhance the success probability of fusion. As a representative example, we here propose a linear-optic scheme for implementing arbitrary $(n,m)$-generalized Shor code or parity code~\cite{Ralph05}. Dual-rail qubits are used to define the encoded-qubits as described below:

The logical basis is written by $|0_L\rangle = |+^{(m)}\rangle^{\otimes n} $ and $|1_L\rangle = |-^{(m)}\rangle^{\otimes n} $, where $|\pm^{(m)}\rangle = (|H\rangle^{\otimes m} \pm |V\rangle^{\otimes m})/\sqrt{2}$. Each logical qubit consists of $n$ blocks, each of which is defined in the form of GHZ state with $m$ photons. Interestingly, the logical Bell states in $(m, n)$-Shor code can be decomposed into $n$-number of block-level Bell states, each of which can be in turn decomposed into $m$-number of photonic Bell states \cite{Lee19} as described following. 
In specific, the logical Bell states in $(m, n)$-Shor code, $|\Psi^\pm\rangle = |0_L\rangle|1_L\rangle \pm |1_L\rangle |0_L\rangle $ and $|\Phi^\pm\rangle = |0_L\rangle|0_L\rangle \pm |1_L\rangle |1_L\rangle $ can be rewritten in terms of block-level Bell states,
\begin{eqnarray}
|\Psi^{+(-)}\rangle &=& \frac{1}{\sqrt{2^{n-1}}}\sum_{j = even(odd) \leq n} \mathcal{P}\Big{[}|\psi^-_{(m)} \rangle ^{\otimes j} |\psi^+_{(m)}\rangle ^{\otimes n-j} \Big{]} \nonumber \\ 
|\Phi^{+(-)}\rangle &=& \frac{1}{\sqrt{2^{n-1}}}\sum_{j = even(odd) \leq n} \mathcal{P}\Big{[}|\phi^-_{(m)} \rangle ^{\otimes j} |\phi^+_{(m)}\rangle ^{\otimes n-j} \Big{]} \nonumber,
\end{eqnarray}
where the $|\phi_{(m)}\rangle$ and $|\psi_{(m)}\rangle $ represent block-level Bell states composed of $2m$ photons, and $\mathcal{P}[\cdot]$ is a permutation function. 
Each block-level Bell state, which has the form
\begin{eqnarray*}
|\psi^\pm _{(m)}\rangle &=& (|+^{(m)}\rangle|-^{(m)}\rangle \pm |-^{(m)}\rangle|+^{(m)}\rangle)/\sqrt{2} \\
|\phi^\pm _{(m)}\rangle &=& (|+^{(m)}\rangle|+^{(m)}\rangle \pm |-^{(m)}\rangle|-^{(m)}\rangle)/\sqrt{2},
\end{eqnarray*}
can be also decomposed into photonic Bell states as 
\begin{eqnarray}
|\psi^\pm_{(m)}\rangle &=& \frac{1}{\sqrt{2^{m-1}}}\sum_{k = odd \leq m} \mathcal{P}\Big{[}|\psi^\pm \rangle ^{\otimes k} |\phi^\pm \rangle ^{\otimes m-k} \Big{]} \nonumber \\
|\phi^\pm_{(m)}\rangle &=& \frac{1}{\sqrt{2^{m-1}}}\sum_{k = even \leq m} \mathcal{P}\Big{[}|\psi^\pm \rangle ^{\otimes k} |\phi^\pm \rangle ^{\otimes m-k} \Big{]} \nonumber,
\end{eqnarray}
where $|\psi^\pm\rangle=(|+\rangle|-\rangle \pm |-\rangle|+\rangle)/\sqrt{2}$ and $|\phi^\pm\rangle=(|+\rangle|+\rangle \pm |-\rangle|-\rangle)/\sqrt{2}$. 

\begin{figure}[t]
  \includegraphics[width=3.2in]{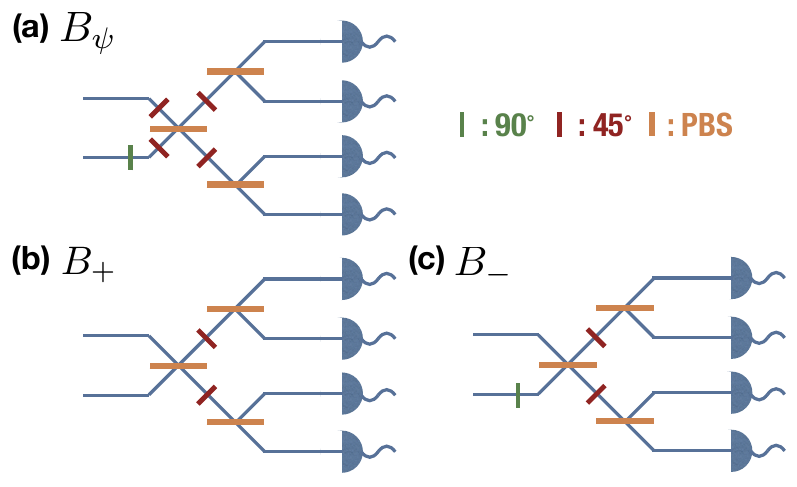}
  \caption{Linear-optical Bell fusion or LOBSM setups for $B_\psi$, $B_+$ and $B_-$. $B_\psi$ can distinguish $|\psi^+\rangle$ and $|\psi^-\rangle $ unambiguously. $B_+$ and $B_-$ discriminate $|\psi^+\rangle$ and $|\phi^+\rangle$, $|\psi^-\rangle$ and $|\phi^-\rangle$, respectively. The Bell states with irrelevance in the BSM setup derive indistinguishable outcomes.}
  \label{BSM}
\end{figure}

Such characteristics of the encoded Bell states allow us to logically distinguish the Bell states by a series of $m \times n$ linear-optic BSMs (LOBSMs) with much higher efficiencies, although each linear-optic BSM in photonic level can distinguish only two out of the four Bell states. We use three different types of linear-optic BSMs discriminating  $|\psi^+\rangle / |\psi^-\rangle$, $|\psi^+\rangle / |\phi^+\rangle$ and $|\psi^-\rangle / |\phi^-\rangle$ deterministically, which are denoted as $B_\psi$, $B_+$ and $B_-$, respectively, and their corresponding setups are illustrated in Fig.~\ref{BSM}. Note that BSM can be implemented by basic linear-optical elements such as polarizing beam splitter (PBS), wave plates and photon detectors, which can discriminate only two out of the four Bell states. The three types can be easily modified by simply rotating wave plates on inputs of PBS. 
Let us explain our encoded-fusion protocol below: 

\begin{enumerate}
\item[(i)] In each block level, we apply $B_\psi$ on each pair of photons randomly selected from distinct encoded-qubits. Repeat until $B_{\psi}$ succeeds, detects a loss, or consecutively fails $j\leq m-1$-times. Here we note that $j$ is a predetermined optimized number for a given loss rate encoding number $(n,m)$.
\item[(ii)] Then, $B_{+}$ or $B_{-}$ is applied on the remaining photon pairs, if  at the step (i) any $ B_{\psi}$ succeeded with the result $\ket{\psi^+}$ and $\ket{\psi^-}$, respectively. For loss detection and $j$-times failure at the step (i), $B_{+}$ or $B_{-}$ is randomly selected and applied on the remaining photon pairs. 
\item[(iii)] Total $n$-times of block-level protocols, (i) and (ii), are performed independently. Then the outcomes are used to discriminate the logical Bell states.
\end{enumerate}

\begin{figure}[t]
  \includegraphics[width=3.0in]{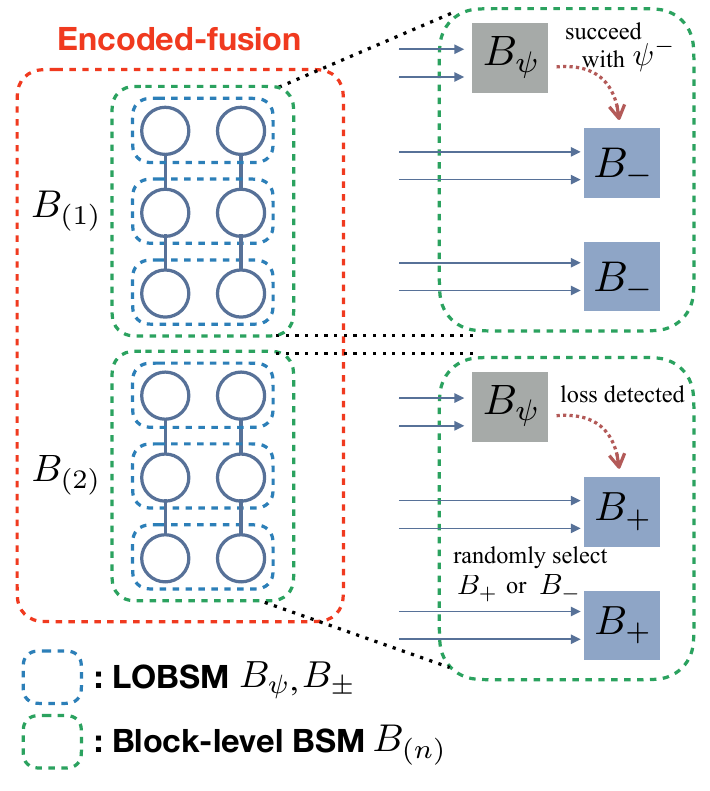}
  \caption{Schematics of encoded-fusion protocol for $(n,m)=(2,3)$ case. An encoded-fusion consists of $n$ block-level BSMs, each of which is composed of $m$ LOBSMs applied according to the result of previous one, i.e., with feed-forwards.}
  \label{EF}
\end{figure}

The outcome of each block is determined based on the outcomes of $m$-times of photonic level LOBSMs: the sign ($\pm$) can be identified by any success of $B_{\psi}$, and the letter ($\psi,\phi$) can be  also identified based on the results of all $B_{\pm}$ performed on remaining photon pairs. Therefore, full discrimination, i.e., both the sign and letter discrimination of the block level Bells states is possible based on all the results of LOBSM unless loss occurs. Moreover, the sign can be identified at least with any single success of $ B_{\psi}$ or $B_{\pm}$ even in the presence of loss. Otherwise, it fails. With the same $\eta$, i.e.,~the same loss rate $1-\eta$ for individual photon, we can write the full discrimination probability of each block level as
\begin{equation}
	p_s(\eta) = \big{(} 1-2^{-j-1} \big{)}\eta^{2m},
    \label{FSuccess}
\end{equation}
which is obtained by excluding the cases of failure without any loss of $m$ photons, i.e., when $B_\psi$ fails $j$-times, and subsequently the random selection of $B_\pm$ turns out to be wrong choice. 
The failure probability can be calculated as
\begin{equation}
	p_f(\eta) = \sum^m _{l=m-j} \Big{(} \frac{1}{2}\Big{)}^{m-l} \eta^{2(m-l)} (1-\eta^2)^l,
    \label{Ffail}
\end{equation}
which represents all possible failure cases, i.e., when all $j$-times of $B_\psi$ fail until loss is detected first, and then subsequently all $B_\pm$ performed on remaining photons detect losses, where $l$ indicates the number of LOBSM where photon loss occurs. 
Then, the probability that we can discrimination the sign only is given as $1-p_s(\eta) -p_f(\eta)$.

\begin{figure}[b]
  \includegraphics[width=3.3in]{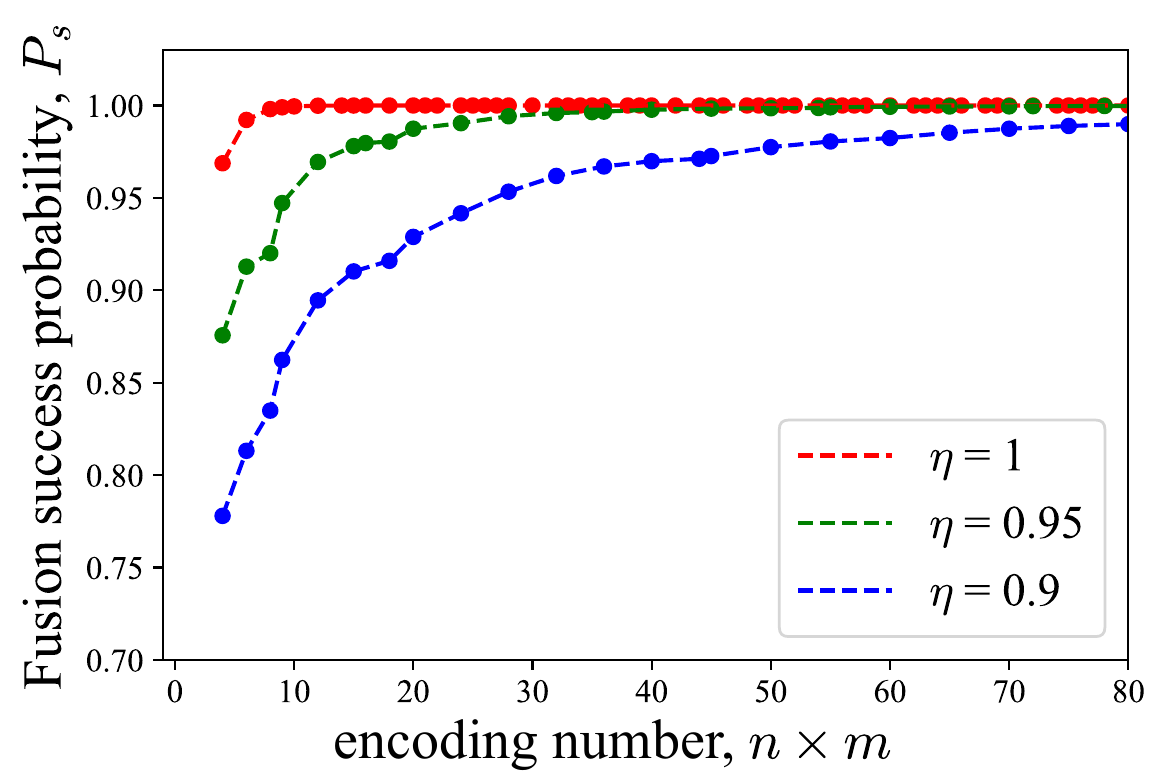}
  \caption{The fusion success probability achieved by our encode-fusion scheme by increasing the encoding number (i.e.~photon number used for encoding) for different $\eta$ (i.e.~loss rate $1-\eta$). The left first points are obtained with encoding $(n,m)=(2,2)$.}
  \label{Psenc}
\end{figure}

By collecting all the outcomes of independently performed $n$-times block-level protocols, the logical result is determined. The letter of logical level is identified as the same one with the letter determined in any block-level result. The sign in logical level can be identified by counting the total number of minus ($-$) sign from block-level outcomes. In the logical level, it is possible to discriminate the Bell states when at least one full-discrimination outcome is obtained and no failure occurs in all $n$-block level protocols. Therefore, the success probability of discriminating the logical Bell states is obtained as 
\begin{equation}
	P_s(\eta) = (1-p_f)^n - (1-p_s -p_f)^n,
\end{equation}
which is plotted in Fig~\ref{Psenc} by increasing the total number of encoded photon in $(n,m)$-Shor code for a given $\eta$. We can see that our Bell-fusion protocol allows us to achieve arbitrarily high success probability by increasing the encoding number $(n,m)$ even under photon losses. It shows that up to unit success probability can be reached under photon losses with a moderate encoding number $(n,m)$. Note that the success probability becomes $1-2^{-mn}$ when no loss occurs. This is in stark contrast to the boost scheme with ancillary entangled photons \cite{Grice11}, by which the success probability has been turned out to be in trade-off with the loss-tolerance.

\section{Resource states}
\label{a:RS}

In this section, we review and study the resource states used for FBQC and EFBQC. While a large-sized entangled resource states is necessary for MBQC, both FBQC and EFBQC require only fixed-sized entangled resource states for computation. Here, we focus on 4-star and 6-ring resources states as described in previous work~\cite{Bartolucci23, Bombin21_A, Sahay23}, both of which are used in fusion networks to construct a RHG lattice implementing surface code. We then consider their extended versions encoded by $(m.n)$-generalized Shor code for EFBQC, called encoded-4-star and encoded-6-ring resource states, respectively. Note that such encoded-resource states are also used for FBQC as introduced in Ref.~\cite{Bartolucci23}. 
We note that the size of the resource states are independent of the computation protocol or code distance to be used in a fusion network. A resource state can be generated by a fixed number of operations from elementary resource states such as Bell states or 3-GHZ states of entangled photons. So, the errors in resource states are bounded. 

Such resource states used here are stabilizer resource states, each of which can be represented by a graph state $|G\rangle$. A graph state can be obtained by preparing qubits in $|+\rangle$ state at each vertex, and then applying controlled-$Z$ gates between adjacent vertices. A graph state $|G\rangle$ corresponds to a stabilizer generator $X_i \prod_{j\in N(i)}Z_j$, where $N(i)$ denotes the set of vertices neighboring vertex $i$, and $X_i$ and $Z_i$ are the Pauli operators acting on qubit $i$.

\begin{figure}[t]
  \includegraphics[width=2.8in]{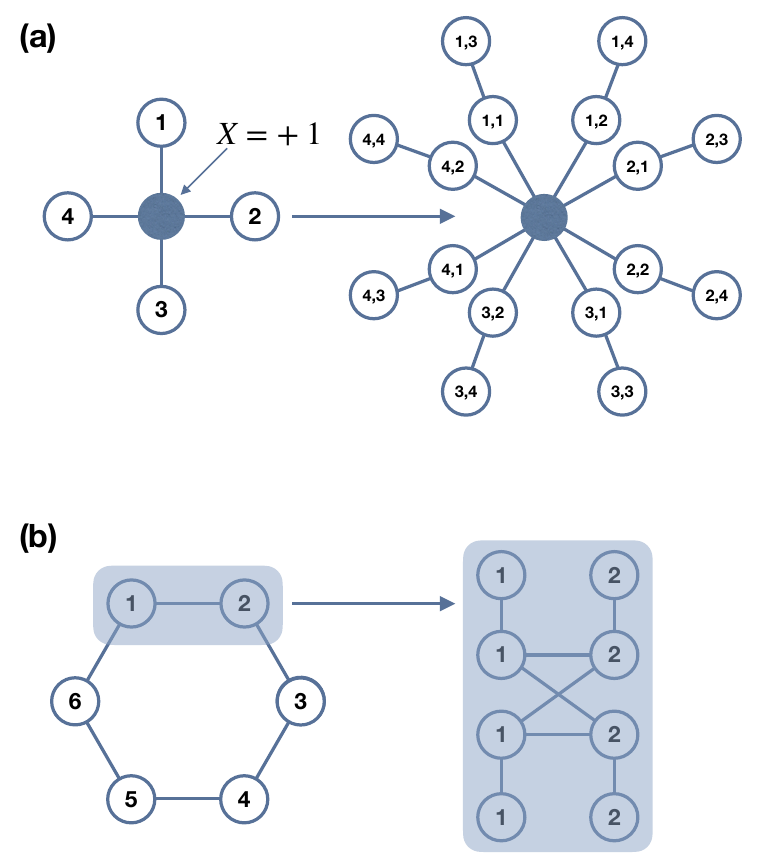}
  \caption{The encoded-resource states (a) 4-star and (b) 6-ring encoded-resource states with $(n,m)=(2,2)$-Shor code. As $n=2$, each branch is duplicated, and as $m=2$, 2 photonic qubits reside in each branch. Both encoded-resource states can be generated by fusing a number of 3-GHZ states. These states are also described to be used for FBQC in Ref.~\cite{Bartolucci23}}
  \label{ERS}
\end{figure}

The 4-star resource state is a four-qubit GHZ state that corresponds to the stabilizer generators $\langle Z_1Z_2Z_3Z_4, X_1X_2, X_2X_3, X_3X_4\rangle $. This state can be also represented as a 5-qubit star graph state with a center qubit measured in the $X$-basis to yield an outcome $+1$. The 6-ring resource state is a six-qubit state, which has a cyclic-form and corresponds to the stabilizer generator $\langle Z_1X_2Z_3,Z_2X_3Z_4,Z_3X_4Z_5,Z_4X_5Z_6,Z_5X_6Z_1,Z_6X_1Z_2\rangle $. 
These resource states can be generated straightforwardly from elementary resource states (e.g.,~3-GHZ states) by using LOBSM or equivalently the type-II fusion~\cite{Browne05, Mercedes15, Li15, Varnava08_A}. A scheme to generate 6-ring resource states using three 3-GHZ states and type-I fusion was also proposed in Ref.~\cite{Sahay23}. 

\begin{figure}[t]
  \includegraphics[width=2.7in]{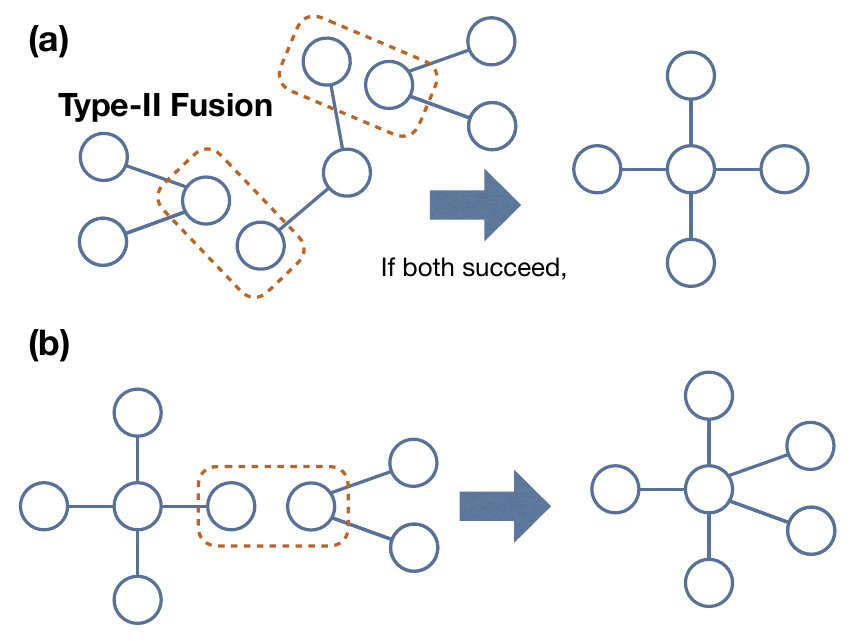}
  \caption{Examples of generation scheme of (a) 5-star graph (5-GHZ) state from three 3-GHZ states via type-II fusions, and (b) $(n+1)$-GHZ state by fusing one dangling edge of $n$-GHZ state and one edge having two dangling edges of 3-GHZ state.}
  \label{RSgen}
\end{figure}
 
\begin{figure}[t]
  \includegraphics[width=2.3in]{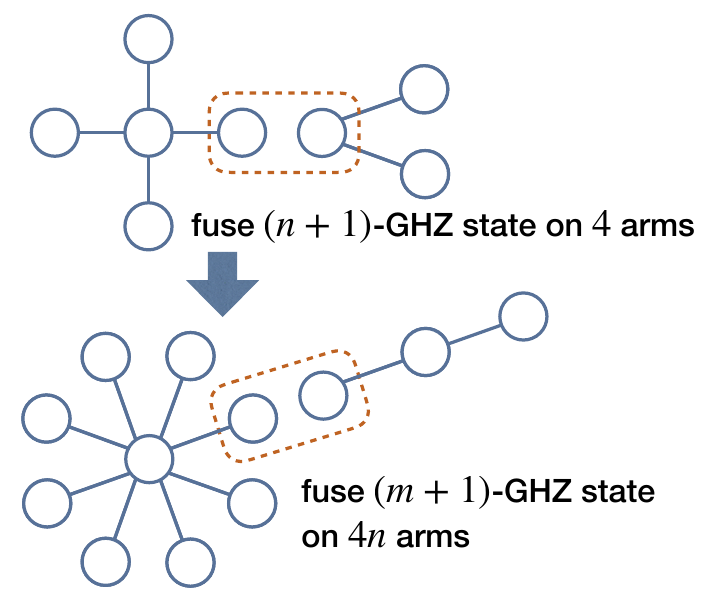}
  \caption{A generation scheme of encoded 4-star resource states in $(n,m)$-Shor code by fusing $4n$-GHZ state and $4 \times n$ number of $(m+1)$-GHZ states.}
  \label{ERSG4}
\end{figure}

\begin{figure}[t]
  \includegraphics[width=2.2in]{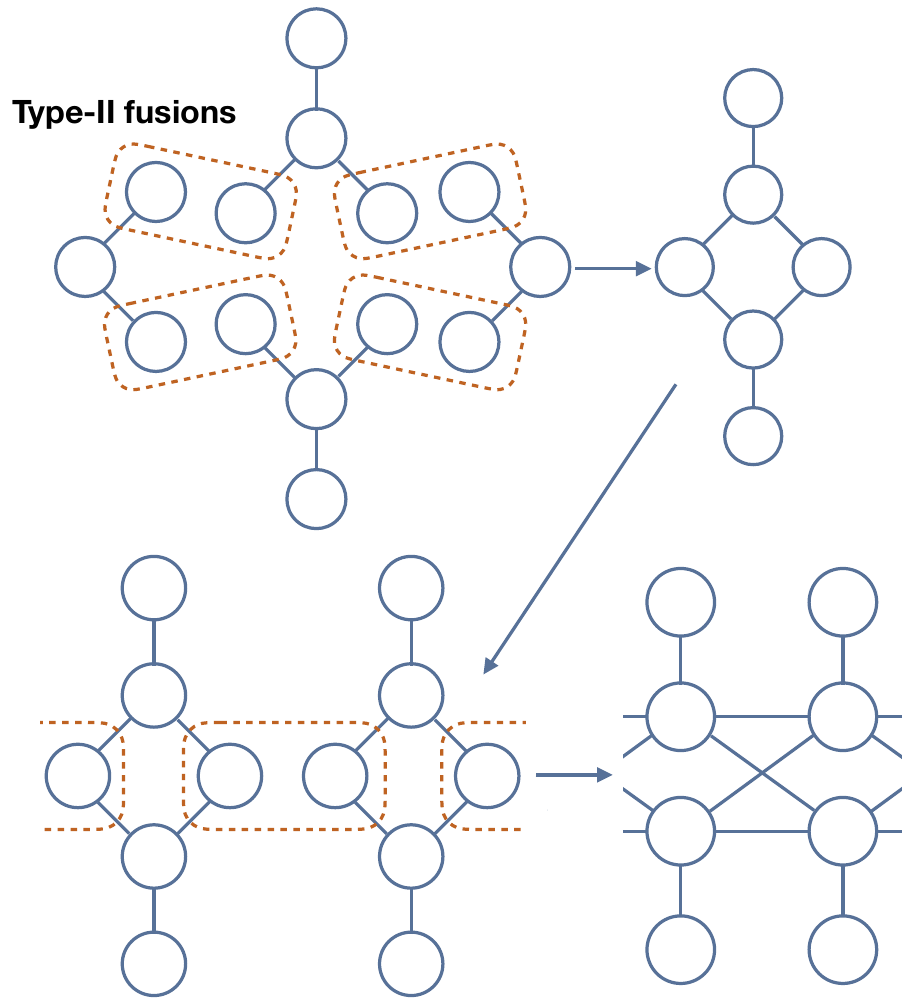}
  \caption{A scheme for generating a $(2,2)$-parity-encoded 6-ring resource state by applying type-II fusion on several 3-GHZ states.}
  \label{ERSG6}
\end{figure}

For the encoded versions with $(n,m)$-Shor code, the qubits participating in the fusion measurements are replaced with the encoded qubits containing $m \times n$ photons. As simplest examples, $(2,2)$-encoded resource states are illustrated in Fig.~\ref{ERS}.
An arbitrary $(n,m)$-encoded 4-star resource state can be generated by fusing several 3-GHZ states as follows. First, arbitrary $n$-GHZ states are constructed from multiple 3-GHZ states as illustrated in Fig.~\ref{RSgen}. Then, an encoded 4-star resource states in $(n,m)$-Shor code can be generated by fusing $4n$-GHZ state and $4 \times n$ number of $(m+1)$-GHZ states as illustrated in Fig.~\ref{ERSG4}. An encoded 6-ring resource state is obtained by replacing  each qubit with encoded qubits in $(n,m)$. Specifically, each qubit is replicated to $n$ copies of qubits, each of which has ($m-1$)-dangling edges. As a simplest example, $(2,2)$-encoded 6-ring resource state can be generated by fusing a number of 3-GHZ states as shown in Fig.~\ref{ERSG6}. 

Recently, the work done by Lee et al.~\cite{Lee23_A} provides an optimal method for generating resource states required for an encoded-fusion based fusion network. They also outlined the required number of elementary resource states, e.g.,~3-GHZ state and the expected number of fusions for generating the target resource state. and we can also use the python package \textit{OptGraphState} proposed in Ref.~\cite{Lee23_A} to estimate the required resources.

\section{Analysis of photon loss threshold in EFBQC}
\label{a:Ana}
EFBQC is performed in the same logical geometry with FBQC, where the qubits and physical fusions are replaced by the encoded-qubits and -fusions, respectively. The noise thresholds of EFBQC can be estimated by a Monte Carlo simulation based on the same hardware-agnostic error model of FBQC analyzed in Ref.~\cite{Bartolucci23}, in which each individual measurement outcome is independently erased with $P_{\text{erasure}}$ and reversed with $P_{\text{error}}$. Here, $P_{\text{erasure}}$ denotes the probability of erasures, i.e.,~the events when the outcome of $XX$ or $ZZ$ is erased. $P_{\text{error}}$ indicates the probability of the event that the fusion outcome is flipped. The correctable region for $P_{\text{erasure}}$ and $P_{\text{error}}$ has been found for 4-star and 6-ring fusion networks. For example, the tolerable $P_{\text{erasure}}$ is 6.90\% for 4-star fusion network and 11.98\% for 6-ring fusion network, if no measurement errors are considered, i.e.~$P_{\text{error}}$ = 0, as found in Ref.~\cite{Bartolucci23}.

The fusion erasure probability $P_{\text{erasure}}$ is determined by two cases: (i) a loss that occurs in fusion process leads to the complete erasure of $XX$ and/or $ZZ$ measurement outcomes, and (ii) a failure of fusion removes either $XX$ or $ZZ$ outcome. By this, the photon loss thresholds can be estimated from the fusion success probability for a given setup and loss rate by checking whether their erasure rates are within the correctable region or not. The results obtained by changing the encoding number for two fusion networks are plotted in Fig.3 in the main text.
These results were obtained from the same fusion erasure thresholds estimated for FBQC, but the properties of fusion and qubits has been modified to the values obtained from the encoded scheme. Within the fusion erasure thresholds, i.e.,~11.98\% for encoded-6-ring and 6.90\% for -4-star fusion networks, correctable regions of fusion success probability and photon loss rates can be in turn estimated based on our encoded-fusion schemes.
It is noteworthy that the fusion success probability and photon loss rate in the existing FBQC are related by a quadratic function, which is because as the number of ancillary entangled photons to boost the fusion success probability increases~\cite{Grice11}, the risk of photon loss also gets higher. This is because any single photon loss while boosting with more photons is leading to the fusion erasure, and as a result the loss threshold becomes lower. On the other hand, in a fault-tolerant fusion network for EFBQC, the number of photons used in fusion increases, the loss tolerance increases simultaneously, so that such a quadratic tendency does not appear.

\begin{figure}[h]
    \includegraphics[width=3.4in]{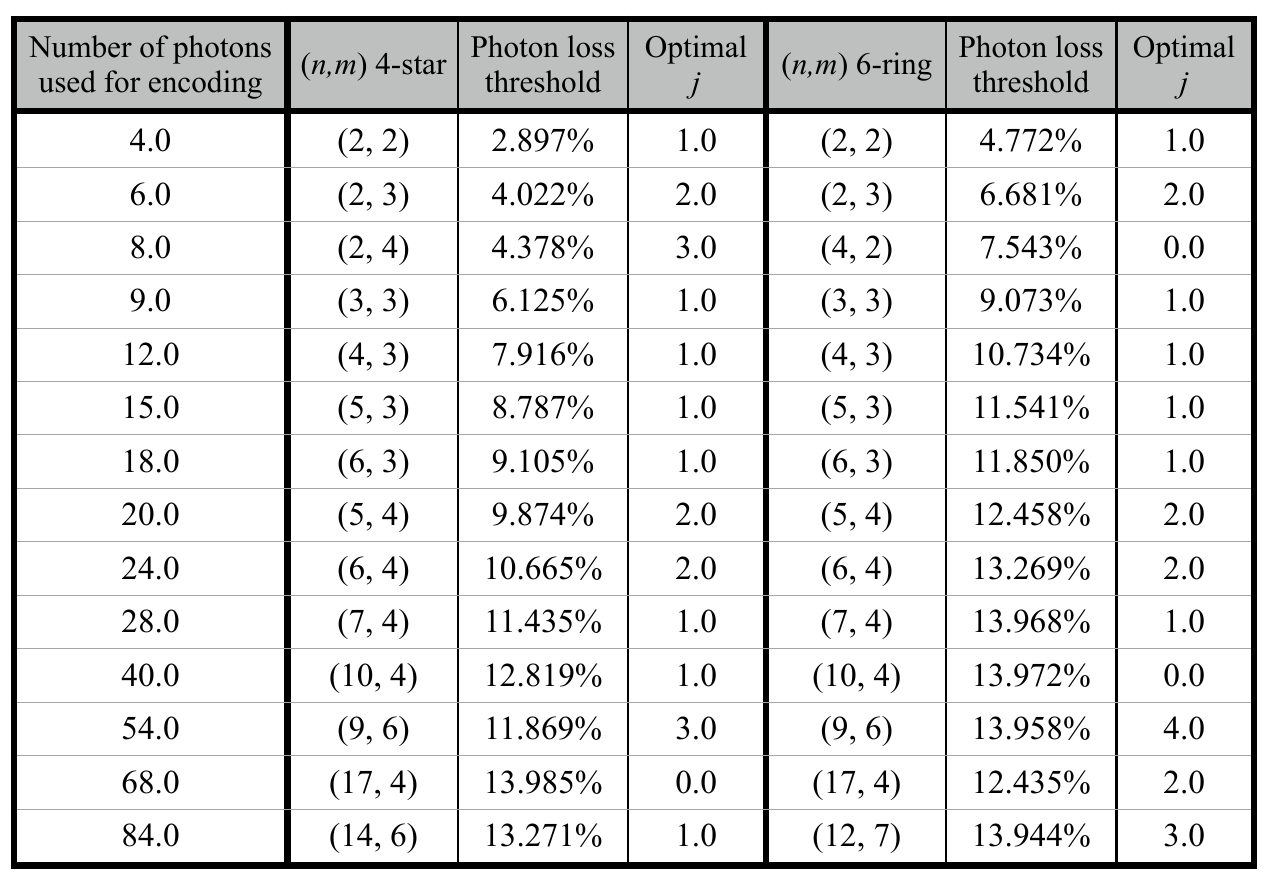}
    \caption{The table for the results of photon loss threshold under various number of photons per fusion. The pairs of $(n,m)$ provide the combinations of $(n,m)$ at a given photon number $n\times m$, to obtain the best value of photons loss threshold together with the optimal parameter $j$.}
    \label{Table_jnm}
\end{figure}

In order to maximize the loss thresholds, the encoded-fusion scheme can be optimized for a given encoding number $(n,m)$ by an additional parameter $j$ which determines the encoded-fusion protocol. In particular, as we can see in Eq.~\ref{FSuccess} and Eq.~\ref{Ffail}, both the block-level success and failure rate are dependent on the parameter $j$. So, $j$ can be optimized for a given $\eta$ and $(n,m)$. Note that the optimized $j$ for the case without loss is always given as $j=m-1$, but in the presence of noise to maximize the loss tolerance, $j$ should be optimized for given other parameters. Importantly, the parameter $j$ reveals the maximum steps of feed-forwards performed in the encoded-fusion protocol, determining the number of steps trying $B_\psi$ with consecutive failure. In other words, we can stop the trials of $B_\psi$ after $j$-step and apply randomly chosen $B_\pm$ on all the remaining pairs with an additional step. Therefore, the optimized encoded-fusion protocol can be executed by less than $j+1$ steps of feed-forwards. 

In the table of Fig.~\ref{Table_jnm}, we present that maximum loss thresholds of 4-star and 6-ring fusion network in EFBQC and the optimized $j$ by increasing the encoding number $(n,m)$, or equivalently the total number of photons used for encoding $n\times m$. The maximum loss thresholds achieved in our work are plotted and compared with the results of FBQC in Fig.~4 in the main text. We note that $\sim 14\%$ loss threshold per photon can be reached, e.g.,~with $(7,4)$ in 6-ring fusion network, and its optimized protocol with $j=1$ can be performed by linear-optical process with one step feed-forward.

\section{Comparison of resource overhead}
\label{a:RO}
Let us compare the resource overheads of FBQC proposed in Ref.~\cite{Bartolucci23} and EFBQC based on our scheme. In the analysis of EFBQC, we have applied our encoded-fusion scheme to the same structure, i.e.,~RHG lattice and the same encoded resource states, i.e.,~generalized Shor code that were used for FBQC in Ref.~\cite{Bartolucci23}. Therefore, a direct comparison of the cost of photons would be possible including the generation process elaborated in Appendix~\ref{a:RS}. Specifically, if both schemes, EFBQC and FBQC, use the same resource states (e.g.,~4-star or 6-ring) with the same encoding size $(n,m)$, a direct comparison is possible irrespective of the generation schemes from any element entangled photons (e.g.,~3-GHZ states).

The maximum threshold 2.7\% of FBQC in Ref.~\cite{Bartolucci23} is achieved with $(2,2)$-encoded 6-ring states plus additional ancillary entangled photons for boosting ($2\times n\times m=2\times2\times2=8$ photons per fusion), while EFBQC by our scheme reaches 4.8\% using the same $(2,2)$-encoded 6-ring states without necessitating ancillary entangled photons. As a result, it is straightforward to see that the overall cost of photons to achieve 2.7\% in FBQC is larger than the overall cost of photons to achieve 4.8\% in EFBQC based on our scheme.

The resource overheads of different schemes toward fault-tolerance are typically compared by estimating the costs to achieve the same threshold. Note that the maximum threshold of FBQC presented in Ref.~\cite{Bartolucci23} is 2.7\%  with $(2,2)$-encoded 6-ring states, which is lower than the minimum of EFBQC 4.8\% with $(2,2)$-encoded 6-ring states as shown in Fig.3 in the main text. A direct comparison of arbitrary encoding size $(n,m)$ may not be thus possible with current data of FBQC in Ref.~\cite{Bartolucci23}. However, from the fact that the loss tolerance of FBQC decays fast by adding more photons from the $(2,2)$-encoding case, it can be estimated that EFBQC significantly outperforms FBQC with respect to the attainable thresholds with given overall number of photon costs by enlarging the encoding size $(n,m)$.

\end{document}